\documentclass[aps,twocolumn,pra,superscriptaddress,amsmath,amssymb]{revtex4-1}
\usepackage{dcolumn}
\usepackage{bm}
\usepackage{amsmath}
\usepackage{txfonts}
\usepackage[T1]{fontenc}
\usepackage{xspace}
\usepackage{ulem}
\usepackage{comment}
\usepackage{braket}
\ifx\pdfoutput\undefined
\usepackage[dvipdfmx]{graphicx}
\usepackage[dvipdfmx]{hyperref}
\usepackage[dvipdfmx]{color}
\else
\usepackage{graphicx}
\usepackage{hyperref}
\usepackage{color}
\fi

\newcommand{\red}[1]{{\color{red} #1}}

\begin{document}
\let\emph\textit

\title{
  Majorana Gap Formation in the Anisotropic Kitaev Model
  with Ordered Flux Configuration
}

\author{Akihiro Hashimoto}

\affiliation{
  Department of Physics, Tokyo Institute of Technology,
  Meguro, Tokyo 152-8551, Japan
}

\author{Yuta Murakami}
\affiliation{
  Center for Emergent Matter Science, RIKEN, Wako 351-0198
}

\author{Akihisa Koga}
\affiliation{
  Department of Physics, Tokyo Institute of Technology,
  Meguro, Tokyo 152-8551, Japan
}

\date{\today}
\begin{abstract}

  We study the Kitaev model with direction dependent interactions
  to investigate how the flux configuration and/or the anisotropy in the exchanges affect 
  the Majorana excitations.
  Systematic numerical calculations demonstrate how 
  the anisotropy of the exchange couplings and flux configuration 
  make the Majorana excitation gapped.
  The induced gapped quantum spin liquid states are distinct from 
  the gapped one realized in the large anisotropic limit.
  The nature of gapped states can be explained by the superlattice potential
  due to flux configuration. 
\end{abstract}
\maketitle

\section{Introduction}

\begin{figure}[htb]
  \includegraphics[width=\linewidth]{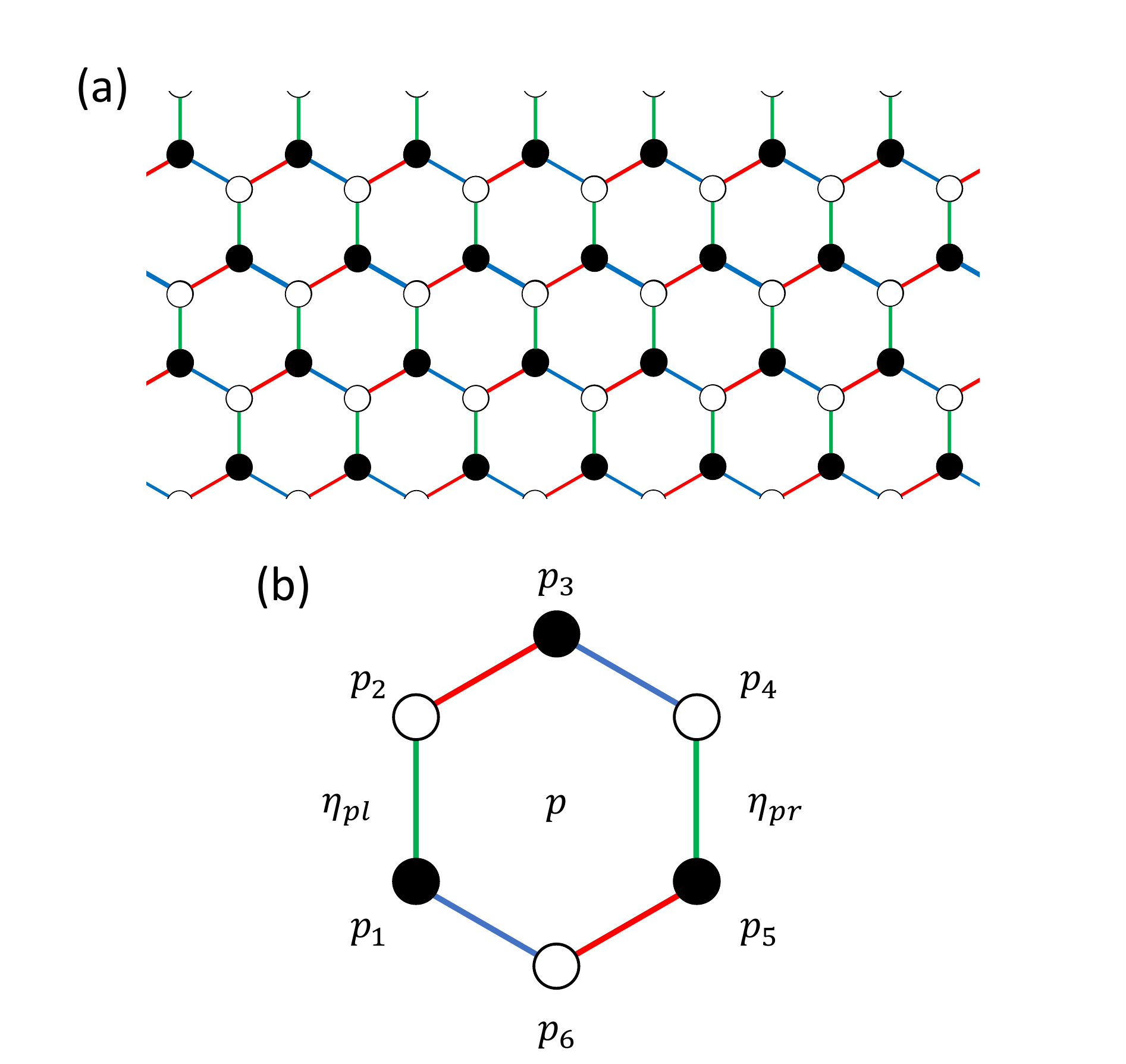}
  \caption{
    (a) Kitaev model with direction dependent Ising exchanges.
    Red, blue, and green lines represent the $x$, $y$ and $z$-bonds.
    (b) Plaquette with sites $p_{1}, p_{2}, \dots, p_{6}$.
    $\eta_{p_l}$ and $\eta_{p_r}$ indicate the local conserved quantities 
    on the $z$-bonds of the left and right edges on the plaquette $p$.
  }
  \label{fig:Kitaev_and_plaquette}
\end{figure}

Last a few decades, quantum spin liquid state has attracted much interest.
One of the interesting examples to realize it is the Kitaev model~\cite{kitaev2006anyons},
whose ground state and finite temperature properties have recently been studied 
in detail~\cite{Jackeli_2009,Chaloupka_2010,Yamaji_2014,Nasu_2014,Nasu_2015,Suzuki_2015,PhysRevB.96.024438,Yamaji_2016,Gohlke_2017,Koga_2018,MotomeNasu2020}.
This model is composed of the direction dependent Ising interactions ($J_x, J_y, J_z$), 
which is schematically shown in Fig.~\ref{fig:Kitaev_and_plaquette}(a).
Due to the existence of the local conserved quantity on each plaquette,
the Kitaev model is solvable and 
the spin degrees of freedom is fractionalized into 
itinerant Majorana fermions and local fluxes.
It is known that, in the ground state, the flux degrees of freedom is frozen
and the quantum spin liquid state is realized.
Low energy properties are then described by the itinerant Majorana fermions.
When the magnitudes of three interactions $J_x, J_y$, and $J_z$ are the same, 
which is called as the "isotropic" case,
the Dirac corn type dispersion appears in the Majorana excitations
and the system is gapless.
Magnetic properties inherent in the Kitaev model have experimentally been examined
in the candidate materials such as
$\rm A_2IrO_3$ (A=Li, Na, Cu)\cite{PhysRevB.82.064412,PhysRevLett.108.127203,PhysRevLett.109.266406,PhysRevLett.108.127204,Kitagawa2018nature,PhysRevLett.114.077202}
and $\alpha$-$\rm RuCl_3$~\cite{Plumb2014,Kubota2015}.
In particular, in the material $\alpha$-$\rm RuCl_3$, 
a half-integer quantized plateau has been observed
in the thermal quantum Hall experiments~\cite{kasahara2018majorana}, 
which should be a direct evidence of a topologically protected
chiral Majorana edge mode. 
As for the bulk properties,
the spin transport mediated by the itinerant Majorana fermions
have theoretically been discussed~\cite{minakawa2020majorana,Taguchi_2021,Taguchi_2022}
although no experiments has been reported so far.

When a certain Ising interaction is much larger than
the others (anisotropic case),
the system should be described by the toric code~\cite{kitaev2003fault}. 
In the case, the quantum spin liquid state is realized
with gapped Majorana excitations and
Majorana correlations exponentially decay.
Therefore, the system can be regarded as a Majorana insulator, 
in contrast to the isotropic case.
It has been claimed that the anisotropy in the exchanges in the candidate materials
can be controlled by the circularly polarized light field~\cite{Arakawa},
which should allow us to control the motion of the Majorana fermions.
Furthermore, the effects of the flux degrees of freedom
on the Majorana excitations have recently been studied~\cite{Czajka,Pereira,Feldmeier,Udagawa,Joy,Nasu2022}.
It has been clarified that the gapped quantum spin liquid state is realized  
when the system has a certain flux configuration~\cite{koga2021majorana}.
It has also been clarified that
this gapped state is not adiabatically connected 
to the gapped one realized in the toric code~\cite{Hashimoto}.
This suggests that the flux configurations and/or anisotropy
in the exchanges should play an important role for
the Majorana excitations in the Kitaev model.
Therefore, it is instructive to clarify
the nature of the gap formation in the Majorana excitation 
in the Kitaev model.

In this paper, we treat the anisotropic Kitaev model on the honeycomb lattice 
to clarify the effects of the triangular flux configurations and/or anisotropy
in the exchange couplings on the Majorana excitations.
Performing systematic calculations, 
we examine how the gap appears in the Majorana excitations.
We clarify that the gapped quantum spin liquid states
are induced by the periodic flux configurations 
and are not adiabatically connected to the gapped state
described by the toric code.

This paper is organized as follows.
In Sec.~\ref{sec:Majorana_and_flux}, 
we introduce the anisotropic Kitaev model and explain our method to treat the flux configuration.
The Majorana gap formation in the system with the triangular flux configuration is discussed in Sec.~\ref{sec:result}.
A summary is given in the last section.



\section{Model and Method}\label{sec:Majorana_and_flux}
We consider the anisotropic Kitaev model, which is described by the following Hamiltonian as
\begin{align}
  H = - J_x \sum_{\langle i,j \rangle_x} S_i^x S_j^x 
  - J_y \sum_{\langle i,j \rangle_y} S_i^y S_j^y   
  - J_z \sum_{\langle i,j \rangle_z} S_i^z S_j^z,  \label{eq:Hamiltonian}
\end{align}
where $\left\langle i,j \right\rangle_\alpha$ stands
for the nearest-neighbor pair 
on $\alpha (= x,y,z)$-bonds.
$S^\alpha_i(=\frac{1}{2}\sigma^\alpha_i)$ is the $\alpha$-component of 
the $S=1/2$ spin operator at the {\it{i}}th site
and $\sigma^\alpha$ is the $\alpha$-component of the Pauli matrix.
$J_\alpha(>0)$ is the ferromagnetic exchange coupling on the $\alpha$-bond.
The model is schematically shown in Fig.~\ref{fig:Kitaev_and_plaquette}(a).
One of the important features of this model is 
the existence of the local conserved quantity~\cite{kitaev2006anyons}.
The local operator $W_p$ on a plaquette $p$ is defined by
$W_p = \sigma^x_{p_1} \sigma^y_{p_2} \sigma^z_{p_3}\sigma^x_{p_4} \sigma^y_{p_5} \sigma^z_{p_6}$,
where $p_i(i = 1, 2, \cdots, 6)$ is the site on plaquette $p$ [see Fig.~\ref{fig:Kitaev_and_plaquette}(b)].
Since $[ W_p, W_{p'}] = 0, [ W_p, H ] = 0$ and $W_p^2 = 1$,
the operator $W_p$ is a local conserved quantity with eigenvalue $w_p=\pm 1$.
Then, each eigenstate of the Kitaev Hamiltonian can be classified
by the subspace with the set of $w_p$.
It is known that the ground state is realized in the subspace with $w_p=1$ 
for each plaquette~\cite{PhysRevLett.73.2158}.
Therefore, one can regard a plaquette with $w_p = -1$ as a flux
and the subspace of the ground state as a flux-free one.

To discuss low energy properties in the Kitaev model,
we use the Jordan-Wigner transformation~\cite{chen2007exact,feng2007topological,chen2008exact} and 
obtain the Hamiltonian in the Majorana representation as,
\begin{align}
  H = &-\frac{i J_x}{4} \sum_{\left\langle rb,r'w \right\rangle_x} \gamma_{rb}\gamma_{r'w}
  -\frac{i J_y}{4} \sum_{\left\langle rb,r'w \right\rangle_y} \gamma_{rb}\gamma_{r'w} \nonumber \\
  &-\frac{i J_z}{4} \sum_{r} \eta_r \gamma_{rb}\gamma_{rw},\label{HH}
\end{align}
where $\gamma_{rb} (\bar{\gamma}_{rw})$ is the itinerant (localized) Majorana fermion operator at the black (white) site 
on the $r$th $z$-bond and $\eta_r = i \bar{\gamma}_{rb} \bar{\gamma}_{rw}$ 
[see Fig.~\ref{fig:Kitaev_and_plaquette}(a)].
Since $[ \eta_r,\eta_{r'} ]=0, [ \eta_r,H ] = 0$ 
and $\eta_r^2 = 1$,
$\eta_r$ is a $Z_2$ local conserved quantity.
It is known that $W_p = \eta_{p_l}\eta_{p_r}$,
where $p_l\; (p_r)$ is the left (right) $z$-bonds on plaquette $p$.
Therefore, the flux configuration $\{w_p\}$ can be represented 
by the configuration $\{\eta_r\}$ instead.
The ground state can be characterized by the subspace with
$\eta_r=1$ for each $z$-bond.

\begin{figure}
  \includegraphics*[width=\linewidth]{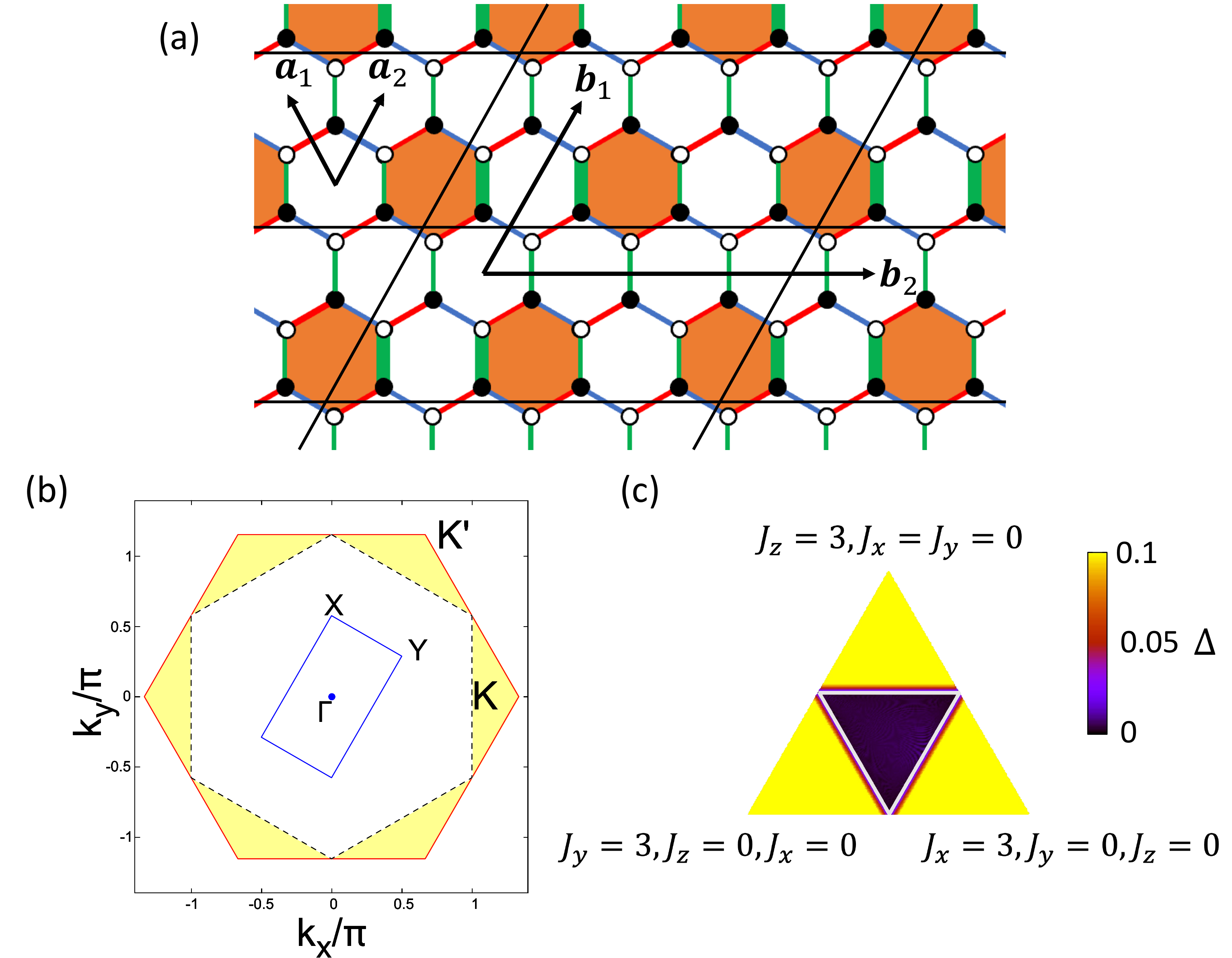}
  \caption{
    (a) The flux configuration ($q=2$) and primitive vectors.
    (b) The red (blue) line shows the original (reduced) Brillouin zone in the Kitaev model.
    Shaded regions show the possible area of the gapless point $\bm{k}_G$ in the flux-free system (see text). 
    (c) Majorana excitation gap in the flux-free system.
    The gapless state appears inside the triangle region specified by the white lines.
  }
  \label{fig:superlattice_and_Brillouin}
\end{figure}

Ground state properties in the flux-free subspace
are well examined~\cite{kitaev2006anyons}.
The dispersion relation for itinerant Majorana fermions is given by
\begin{align}
  \epsilon(\bm{k}) = \frac{1}{2} \Big| J_x \exp(i \bm{k} \cdot \bm{a}_2) + J_y \exp (i\bm{k}\cdot\bm{a}_1) + J_z \Big|,
\end{align}
where $\bm{a}_1$ and $\bm{a}_2$ are primitive vectors of the honeycomb lattice 
[see Fig.~\ref{fig:superlattice_and_Brillouin}(a)],
and $\bm{k}$ is the wave vector.
In the isotropic case with $J_x=J_y=J_z$, 
the linear dispersion appears and
the gapless points are located at K and K' point in the Brillouin zone, 
as shown in Fig.~\ref{fig:superlattice_and_Brillouin}(b).
Introducing the anisotropy in the exchange,
the gapless points gradually change.
When the set of exchanges satisfies triangle inequalities 
[triangle region in the diagram shown in Fig.~\ref{fig:superlattice_and_Brillouin}(c)], 
\begin{align}
  J_x + J_y &\geq J_z, \\
  J_y+J_z &\geq J_x, \\
  J_z + J_x &\geq J_y,
\end{align} 
the system is gapless, and
the gapless points take the inside of a certain region in the Brillouin zone 
shown as a shaded area in Fig.~\ref{fig:superlattice_and_Brillouin}(b).
On the other hand,
when the exchanges are away from the triangle inequalities,
the excitation gap appears in the Majorana excitation.
In the case, the system should be adiabatically connected to that for the dimer limit ($J_x=J_y=0$),
and thereby low energy properties are effectively described by the toric code~\cite{kitaev2003fault}.
It is also known that Majorana excitations are controlled
by not only the anisotropy in the exchanges, but also
the flux configurations~\cite{koga2021majorana}.
Therefore, it is necessary to clarify the gap formation in the Majorana systems
and the role of flux configurations and/or anisotropy in the exchange.

In this study, we focus on the triangular flux configurations,
where the fluxes are periodically arranged in the honeycomb sheet, 
as shown in Fig.~\ref{fig:superlattice_and_Brillouin}(a).
This flux configuration is specified by its unit length $q$.
The flux-free configuration corresponds to the limit  $q\rightarrow\infty$.
When the triangular flux configuration is represented by the set of $\{\eta_r\}$,
the unit cell is characterized by the primitive vectors
$\bm{b}_1 =q \bm{a}_2$ and $\bm{b}_2 = -2q\bm{a}_1 + 2q\bm{a}_2$
[see Fig.~\ref{fig:superlattice_and_Brillouin}(a)].
We should note that the unit cell specified by the $\{ \eta_r \}$ is different from the one by the flux.
The set of $\{\eta_r\}$ for the flux configuration $q=2$
is shown as the bold and thin lines on the $z$-bonds.
The Brillouin zone for the original Kitaev model is given by the hexagon in the Fourier space,
and reduced one for flux configurations $q$ 
is given by the rectangle, as shown in Fig.~\ref{fig:superlattice_and_Brillouin}(b).
In the following, we study the Majorana excitation under the conditions $J_x+J_y+J_z = 3$.
We note that, in the system with the triangular flux configuration,
the model Hamiltonian eq.~(\ref{HH}) is symmetric under the exchange operation
in $\{J_x, J_y, J_z\}$.

Before starting with discussions,
we comment on magnetic properties in two limits of the system with the flux configuration $q$.
When $J_z$ is large, the system is reduced to
the weakly-coupled dimers, where the Majorana gap
is $\Delta\sim J_z$ and the flux configuration is irrelevant.
In the case, the effective Hamiltonian should be given by
the fourth-order perturbation theory and
the system is described by the toric code.
When $J_z=0$, the system is reduced to isolated one-dimensional
spin chains composed of $x$ and $y$-bonds.
The system is then described by the free Majorana fermions and
the flux configuration is also irrelevant.
In the following, we discuss how the flux configurations affect low energy Majorana excitations
in the Kitaev model away from these two limits.

\section{Result}\label{sec:result}

\begin{figure}[htb]
  \includegraphics[width=\linewidth]{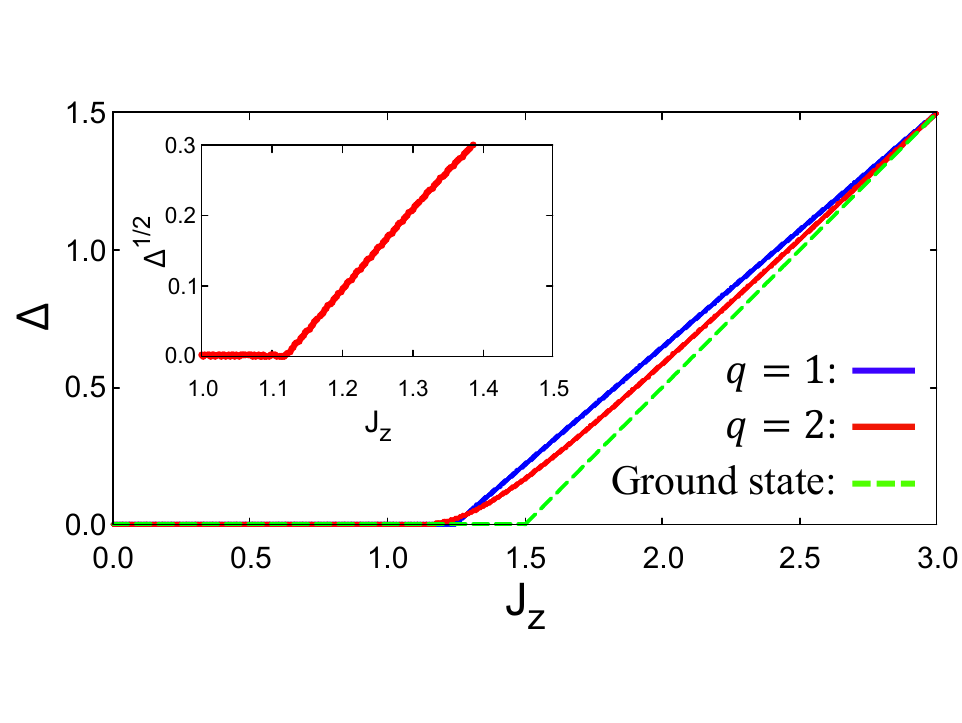}
  \caption{
    Blue and red lines represent Majorana excitation gap $\Delta$ in the systems
    with the flux configurations $q=1$ and $q=2$ under the conditions $J_x+J_y+J_z=3$ and $J_x = J_y$.
    Dashed line represent the results for the flux-free state. 
    The inset shows $\Delta^{1/2}$ as a function of $J_z$ when $q=2$.
  }
  \label{fig:q12Gaps}
\end{figure}

We consider the triangular flux configuration
to discuss Majorana excitations.
Diagonalizing the Hamiltonian with the corresponding configurations $\{\eta_r\}$,
we obtain the Majorana dispersion relations.
First, we consider the flux configuration with $q=1$, where
the system is fully covered by the fluxes.
Now, the Majorana gap is examined under the condition with $J_x=J_y$,
as shown in Fig.~\ref{fig:q12Gaps}.
When $J_z=0$, the system is reduced to one-dimensional chains, where
the flux configuration plays no role
for the Majorana excitation and the system is gapless.
Introducing $J_z$,
the flux configuration affects low energy properties
and the gapless points change.
On the other hand, the gapless excitation remains until a certain value $(J_z)_c\sim1.243$.
Beyond $(J_z)_c$, the excitation gap linearly increases.
In the case, the system should be effectively described by the toric code,
where the flux configuration is irrelevant.
Therefore, the curve of the excitation gap is similar to that
for the flux-free state,
which is shown as the dashed line in Fig.~\ref{fig:q12Gaps}.

\begin{figure}[htb]
  \includegraphics[width=\linewidth]{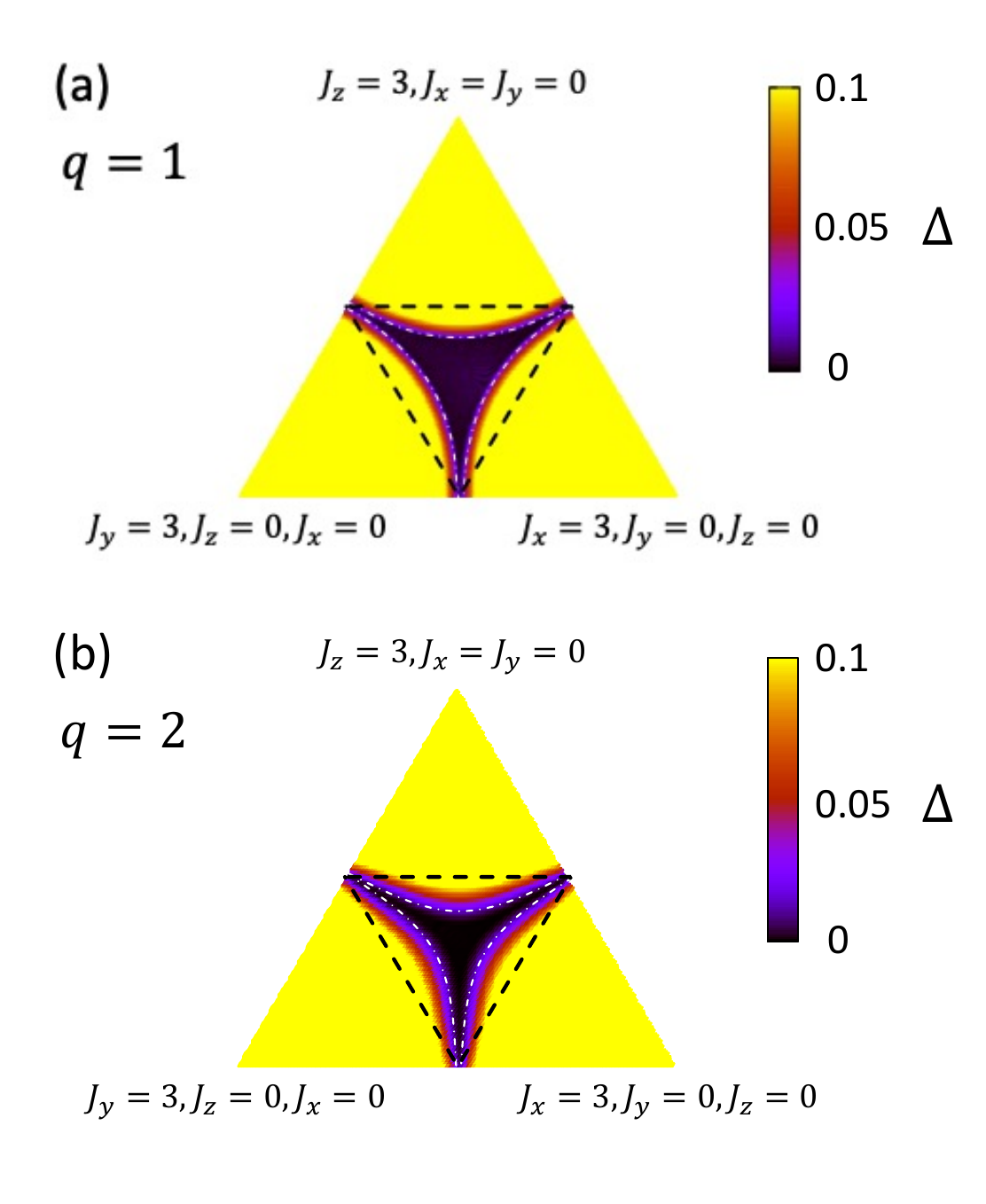}
  \caption{
    Majorana excitation gap in the system with the flux configurations 
    (a) $q=1$ and (b) $q=2$ on the plane $J_x+J_y+J_z=3$.
  }
  \label{fig:q12Phases}
\end{figure}
By performing similar calculations,
we obtain the Majorana excitation gap in the parameter space with $J_x+J_y+J_z=3$, 
as shown in Fig.~\ref{fig:q12Phases}(a).
We find that the gapless ground state is realized 
in the isotropic case $(J_x=J_y=J_z)$~\cite{koga2021majorana}, 
and is stable against the small anisotropy in the exchange couplings.
Similar behavior is also found in the $q=2$ case,
as shown in Fig.~\ref{fig:q12Gaps} and Fig.~\ref{fig:q12Phases}(b).
Therefore, we can say that the flux configurations play a minor role
in the Majorana excitations when $q=1$ and $q=2$.
Some detail of the dispersion relations in the isotropic case 
is discussed in Appendix~\ref{sec:appendix_near_zero_point}.

\begin{figure}[htb]
  \includegraphics[width=\linewidth]{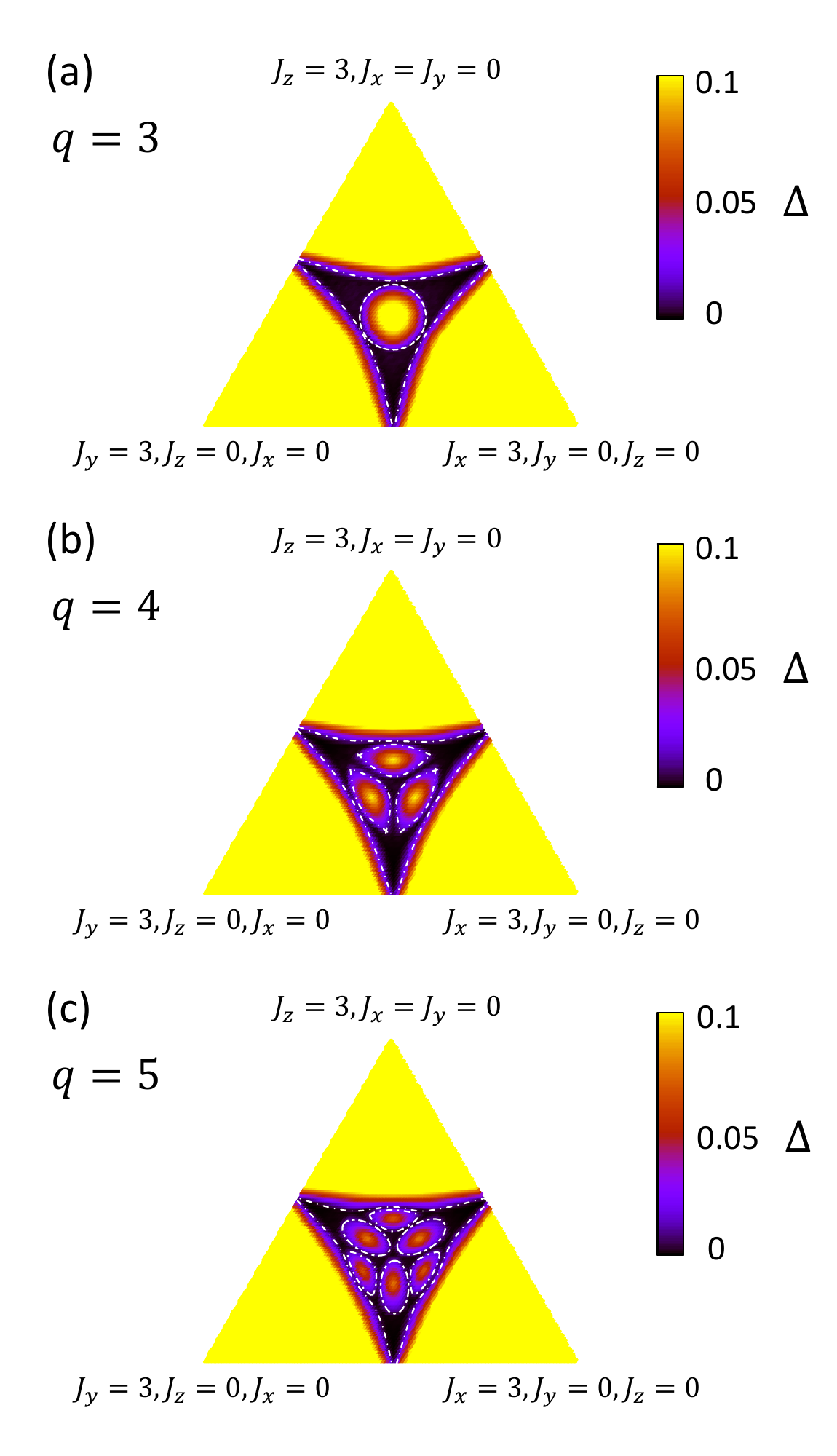}
  \caption{
    Majorana excitation gap in the system with the flux configurations 
    (a) $q=3$, (b) $q=4$, and (c) $q=5$ on the plane $J_x+J_y+J_z=3$.
    }
  \label{fig:q345}
\end{figure}

When $q\ge 3$, distinct behavior appears in the Majorana excitations.
When $q=3$, we find in Fig.~\ref{fig:q345}(a)
the gapped state around the isotropic point.
The crosssection under the condition $J_x=J_y$ clearly indicates
three transition points at $J_z=0.678, 1.308$, and $1.313$, as shown in Fig.~\ref{fig:q345}(d).
This implies that, around the isotropic point, the gapped quantum spin liquid
is driven by the triangular flux configuration.
Furthermore, this gapped region is bounded by the gapless region
although the region is narrow under the condition $J_x=J_y$.
Therefore, we can say that this gapped state is not
adiabatically connected to the gapped one realized in the large $J_z$ limit.
In the gapped state, 
the Majorana excitation gap takes its maximum $\Delta=0.153$
at the isotropic point $J_x=J_y=J_z$.

\begin{figure}[htb]
  \includegraphics*[width=\linewidth]{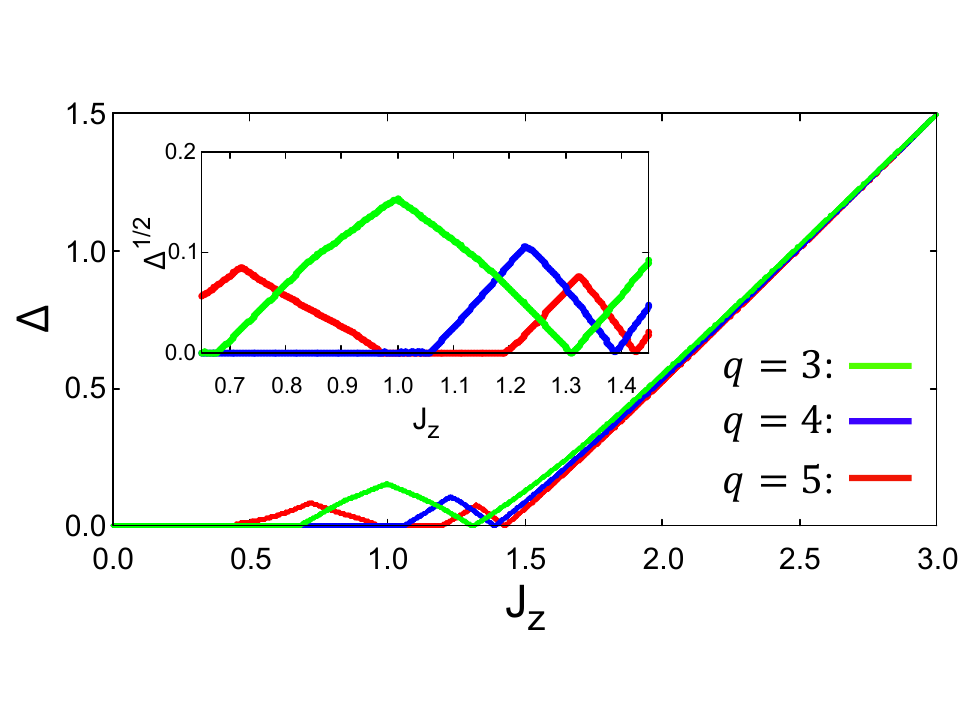}
  \caption{
    Majorana excitation gap $\Delta$ as a function of $J_z$ under the condition $J_x=J_y$ 
    when $q=3$ (green), $q=4$ (blue), and $q=5$ (red).
    Inset is the magnified figure around $J_z=1$. 
    }
  \label{fig:gap_q345}
\end{figure}

The number of the gapped states stabilized by the flux configurations
is one for $q=3$, three for $q=4$, and six for $q=5$,
as shown in Fig.~\ref{fig:q345}.
We note that, for the cases $q=4$ and $q=5$,
the system is gapless in the isotropic point.
This implies that the anisotropy in the exchange interactions
as well as the flux configurations
plays an important role in realizing the gapped states.
The maximum of the gap is located on the axis of $J_x=J_y$
(and its equivalent axes),
and the corresponding exchanges are given as
$J_z\simeq 1.228$ for $q=4$ and $J_z\simeq 0.720, 1.323$ for $q=5$,
as shown in Fig.~\ref{fig:gap_q345}.

\begin{figure}[htb]
  \includegraphics*[width=\linewidth]{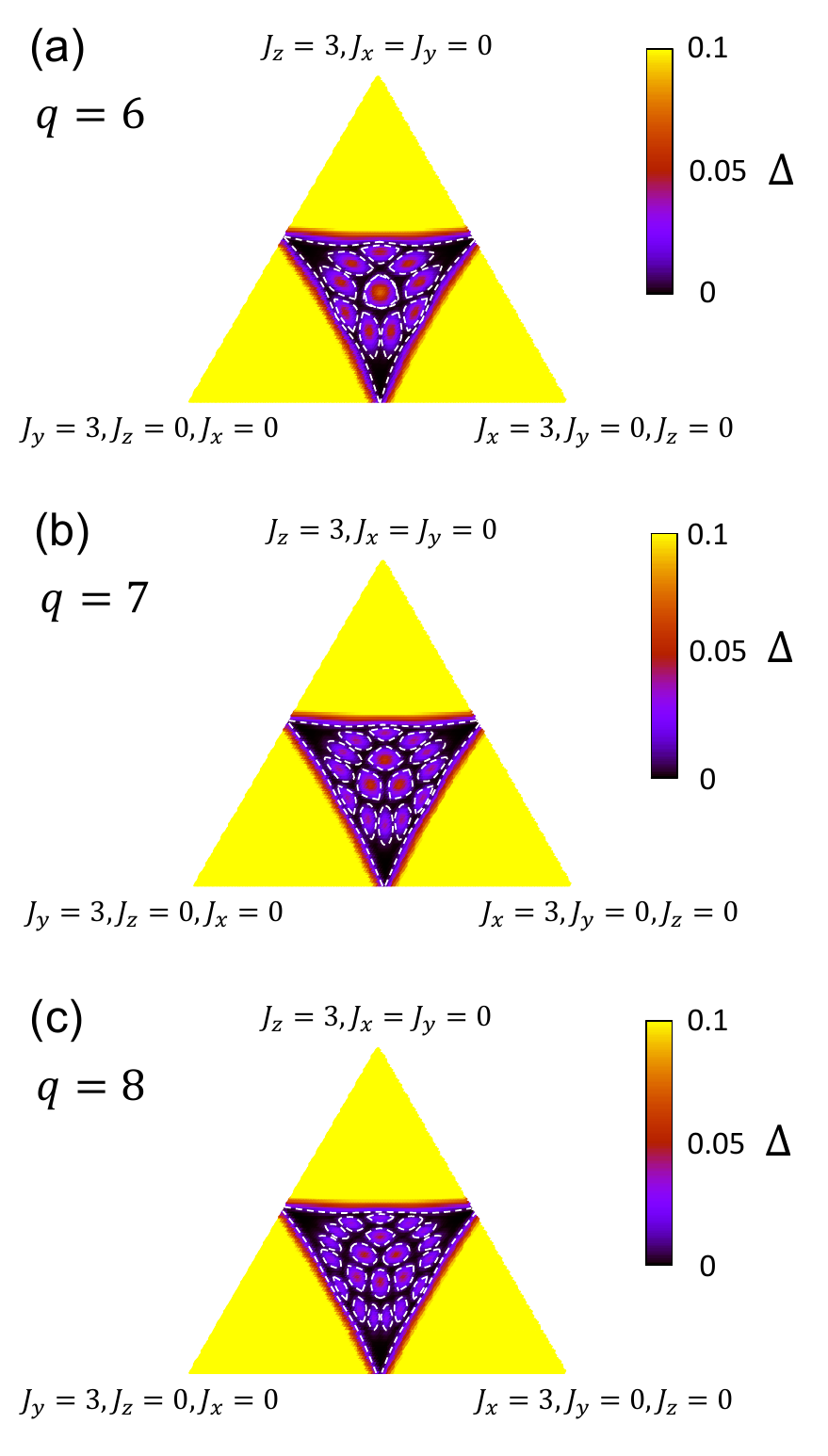}
  \caption[]{
    Majorana excitation gap in systems with flux configurations $q=6, 7$, and $8$.
    }
  \label{fig:phases_q678}
\end{figure}
Figure~\ref{fig:phases_q678} shows the Majorana excitation gap 
in the systems with $q=6, 7$, and 8. 
We find several gapped states in the triangular region.
The number of the gapped states is represented by $(q-1)(q-2)/2$ for the flux configuration $q$.
An important point is that the bilayer structure appears
from the isotropic point $J_x=J_y=J_z$.
In the case with $q=6$,
the gapped state is realized at the isotropic point,
and away from this state (second layer), nine distinct gapped states are realized.
When $q=7$ ($q=8$), three (six) gapped states appear in the first layer,
and twelve (fifteen) gapped states appear in the second layer.
These results should suggest that ``three'' is a key role in the Kitaev system~\cite{koga2021majorana}.
In fact, the increase $q$ by three increments the number of the layer, 
which has been confirmed in the system with the flux configurations with $q\red{\ge} 10$ (not shown).

\begin{figure}
  \includegraphics[width=\linewidth]{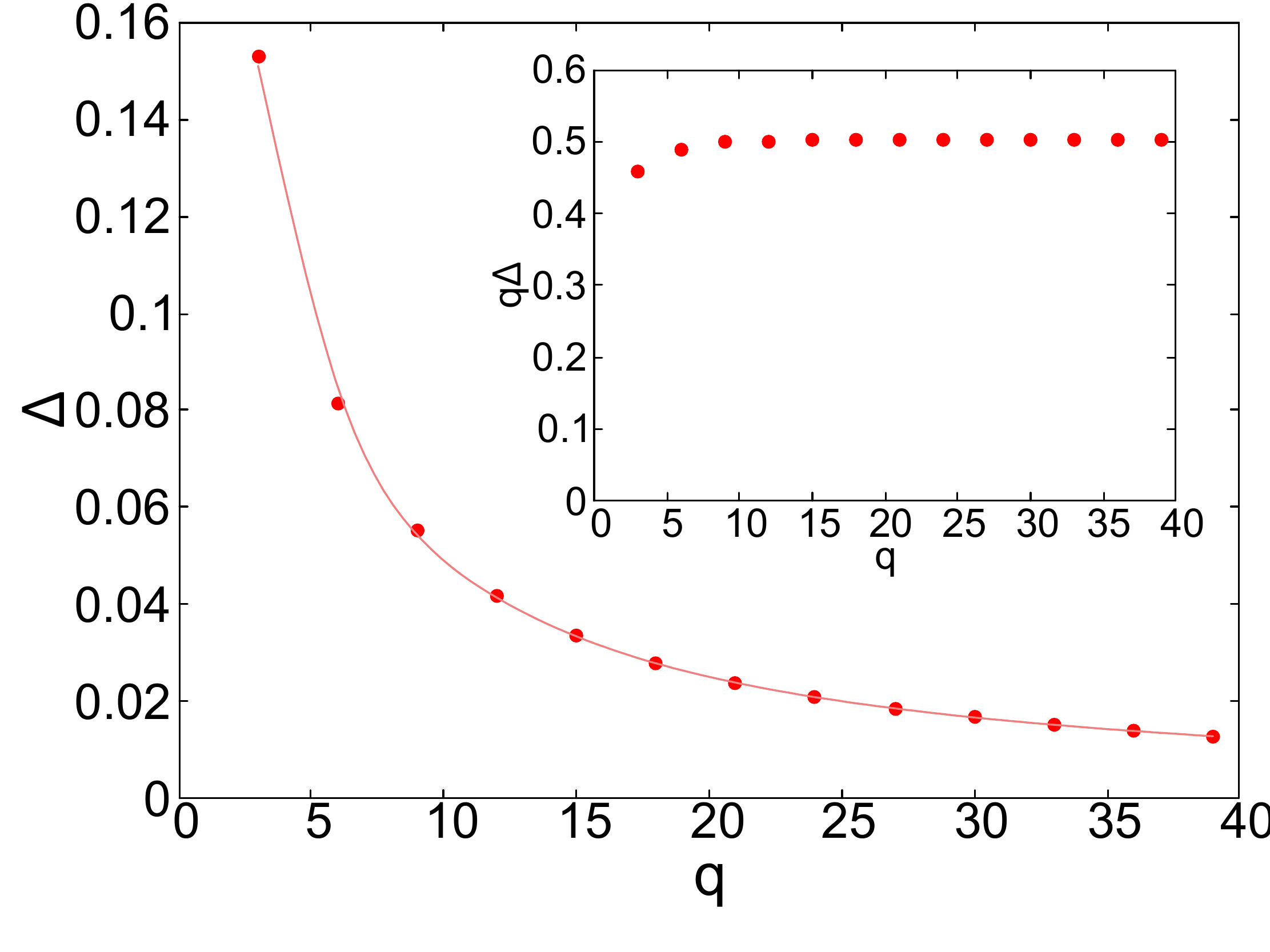}
  \caption{
    Majorana excitation gap $\Delta$ in the system with the flux configuration $q=3n$ with integer $n$.
    The inset shows $q\Delta$ with respect to $q$.   
  }
  \label{fig:gap3n}
\end{figure}

When $q=3n$ with integer $n$,
the Majorana excitation gap appears at the isotropic point, as shown in Fig.~\ref{fig:gap3n}.
Increasing $q$, the Majorana excitation gap monotonically decreases.
Since $q\Delta$ is nearly constant with respect to changes in q,
$\Delta$ is inversely proportional to $q$.
In the case, the numbers of the layers and gapped states increase
in the triangle region
$J_x + J_y \geq J_z, J_y+J_z \geq J_x, J_z + J_x \geq J_y$.
These fact should be consistent with the fact that the gapless ground state is realized there
when $q\rightarrow\infty$.

The Majorana gap formation in the system with the triangular flux configuration may be simply explained, 
by taking into account the idea of the superlattice potential.
In the flux-free case, the gapless point $\bm{k}_G$ satisfies $\epsilon(\bm{k}_G)=0$.
Since $\bm{k}_G\neq 0$, two gapless points are given by $\pm\bm{k}_G$
and are located inside certain regions, 
which are shown as the shaded areas in Fig.~\ref{fig:superlattice_and_Brillouin}(b).
Now, one takes into account the reduced Brillouin zone for the flux configuration, 
where the gapless points $\pm\bm{k}'_G$ are defined in the reduced Brillouin zone.
When $\bm{k}_G'=-\bm{k}_G'+\bm{K}'_G$, where $\bm{K}'_G$ is the reciprocal lattice vector,
the periodic potential for the Majorana fermions from the flux  should yield the hybridization between two branches,
leading to the Majorana excitation gap.
We note that the M and N points in the reduced Brillouin zone are not genuine symmetric points
since we have treated the triangular flux configurations in terms of the set of $\{\eta_r\}$.
Therefore, the gapped quantum spin liquid state should be realized
when $\bm{k}'_G$ is located at high symmetric points $X, Y$, and $\Gamma$.

\begin{figure}
  \includegraphics[width=\linewidth]{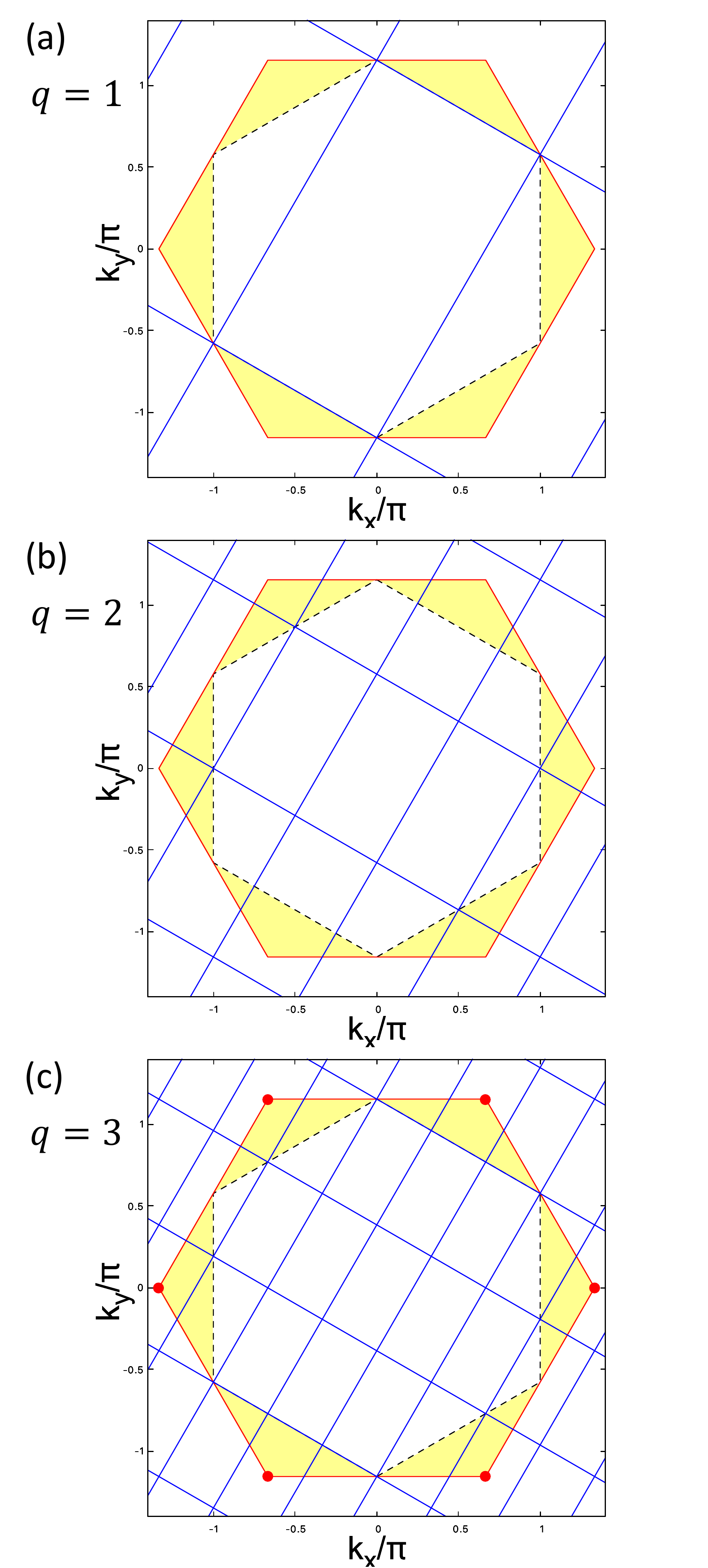}
  \caption{
    Brillouin zones for the flux configurations (a) $q=1$, (b) $q=2$, and (c) $q=3$.   
    Shaded region shows possible area of the gapless point $\bm{k}_G$ in the flux-free system.
    Solid red circles at the symmetric points in the possible area
    are imporant for generating the gapped states (see text).
%
  }
  \label{fig:space123}
\end{figure}
Figure~\ref{fig:space123} shows the reduced Brillouin zones for the 
$q=1$, 2, and 3 cases in the original hexagonal Brillouin zone with 
the shaded area (see also Fig.~\ref{fig:superlattice_and_Brillouin}).
We clearly find there are no high symmetric points
in the corresponding regions for the $q=1$ and 2.
Therefore, the flux configuration plays a minor role in the Majorana excitations,
which is consistent with the absence of the gapped states.
In the case $q=3$,
the region includes the $\Gamma$ point.
Therefore, around the corresponding exchanges $J_x = J_y = J_z$,
the Majorana gap is induced,
which is consistent with the numerical results.

\begin{figure}
  \includegraphics[width=\linewidth]{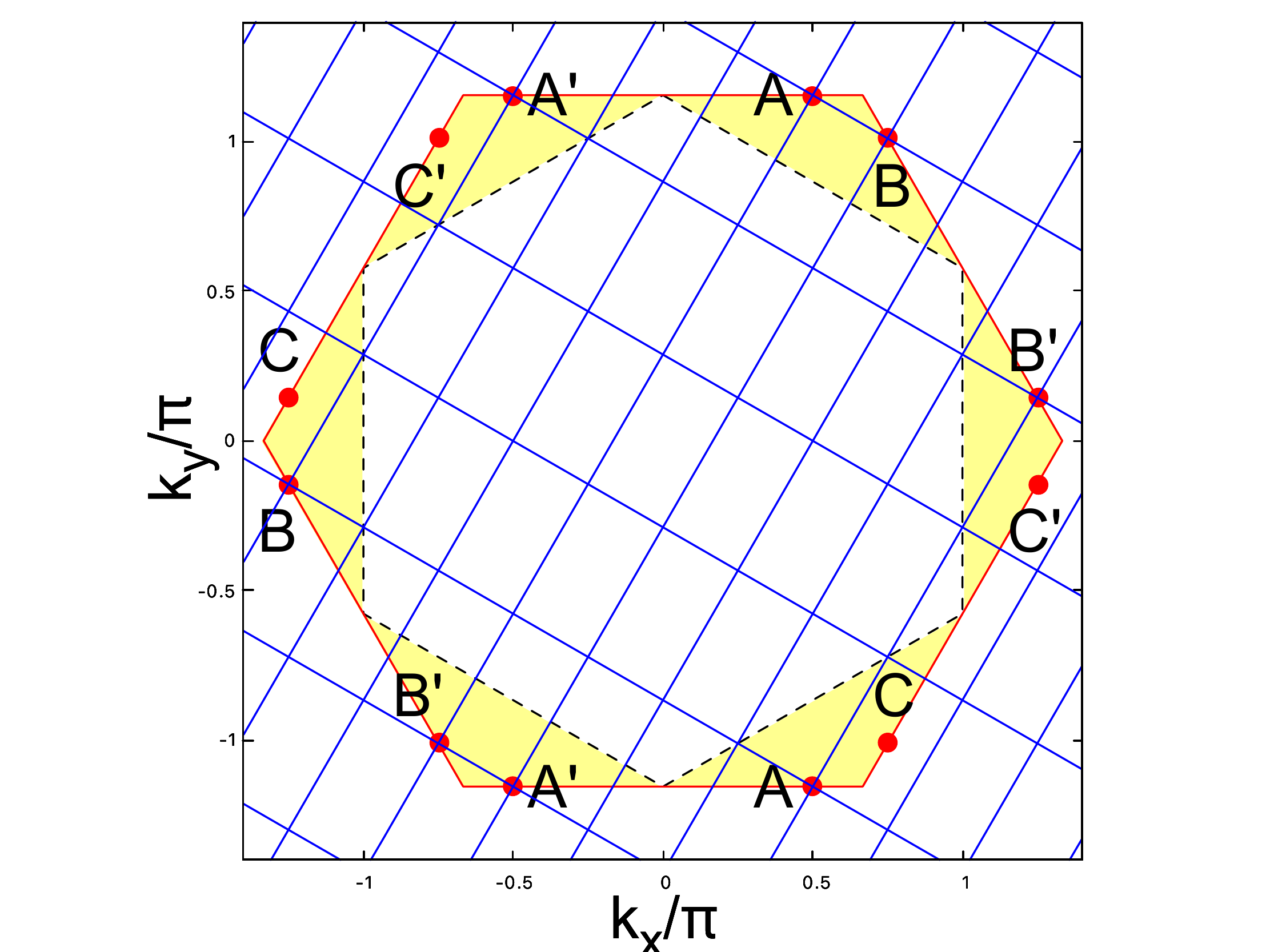}
    \caption{
      Brillouin zones for the flux configuration with $q=4$.   
      Shaded region shows possible area of the gapless point $\bm{k}_G$ in the flux-free system.
      Solid red circles at the symmetric points in the possible area
      are imporant for generating the gapped states (see text).
     }
    \label{fig:space4}
  \end{figure}
Figure~\ref{fig:space4} is the reduced Brillouin zone for the flux configuration $q=4$.
We find three symmetric points in the shaded area.
A (A') point is located at the X point,
B (B') point is at Y point, and
C (C') point is at $\Gamma$ point in the reduced Brillouin zone.
These three points are equivalent under the exchange operation $\{J_x, J_y, J_z\}$. 
The exchange couplings for the A point being the gapless point in the Majorana dispersion
are given $J_x=J_y=3-3/\sqrt{2}$, and $J_z = 3(\sqrt{2}-1)$.
The values are close to $(J_x, J_y, J_z) = (0.886, 0.886, 1.228)$,
where the Majorana excitation takes a maximum in its gapped state.
As for the $q=5$ case,
there are two independent points under the condition $J_x=J_y$.
When $J_z\sim 0.708 (1.342)$,
the gapless point is located $\Gamma$ (X) point in the reduced Brillouin zone.
These are also consistent with the fact that
the Majorana excitation gap takes a maximum in the distinct gapped states.
As $q$ increases, the unit cell of the flux configurations becomes larger 
while the reduced Brillouin zone becomes smaller.
In the case, high symmetry points in the reduced Brillouin zone 
are covered by some shaded areas, 
which leads to the increase of the number of the gapped states.
The number of symmetry points in the shaded area is $(q-1)(q-2)/2$,
which is consistent with the number of the gaped states.
By these reasons, we can say that the triangular flux configuration yields
the periodic potential for the Majorana fermions and 
the excitation gap opens when $\bm{k}_G$ coincides with high symmetry points.
It is naively expected that this can be applied to the Kitaev system
with distinct flux configurations,
which is now under considerations.

\section{Summary}\label{sec:summmary}

We have investigated the anisotropic Kitaev model with triangular flux configurations
to discuss the Majorana excitation.
Systematic numerical calculations have clarified how 
the anisotropy of the exchange couplings and flux configuration 
create the Majorana excitation gap.
The induced gapped quantum spin liquid states are distinct from 
the gapped one realized in the large anisotropic limit.
We have also addressed the $q$-dependence of the gap magnitude of the Majorana excitations. 
These results suggest that Majorana insulators may be realized 
by controlling the anisotropy in the exchanges and/or flux configurations.
We believe that the realization of Majorana insulators will further advance research 
into the development of quantum devices using Majorana fermions.

\begin{acknowledgments}
  Parts of the numerical calculations are performed
  in the supercomputing systems in ISSP, the University of Tokyo.
  This work was supported by Grant-in-Aid for Scientific Research from
  JSPS, KAKENHI Grant Nos. JP20K14412, JP21H05017 (Y.M.), 
  JP22K03525, JP21H01025, JP19H05821 (A.K.), and JST CREST Grant No. JPMJCR1901 (Y.M.).
\end{acknowledgments}

\appendix
\section{The dispersion near the zero point in the isotropic case}\label{sec:appendix_near_zero_point}

\begin{figure*}
  \includegraphics[width=0.8\linewidth]{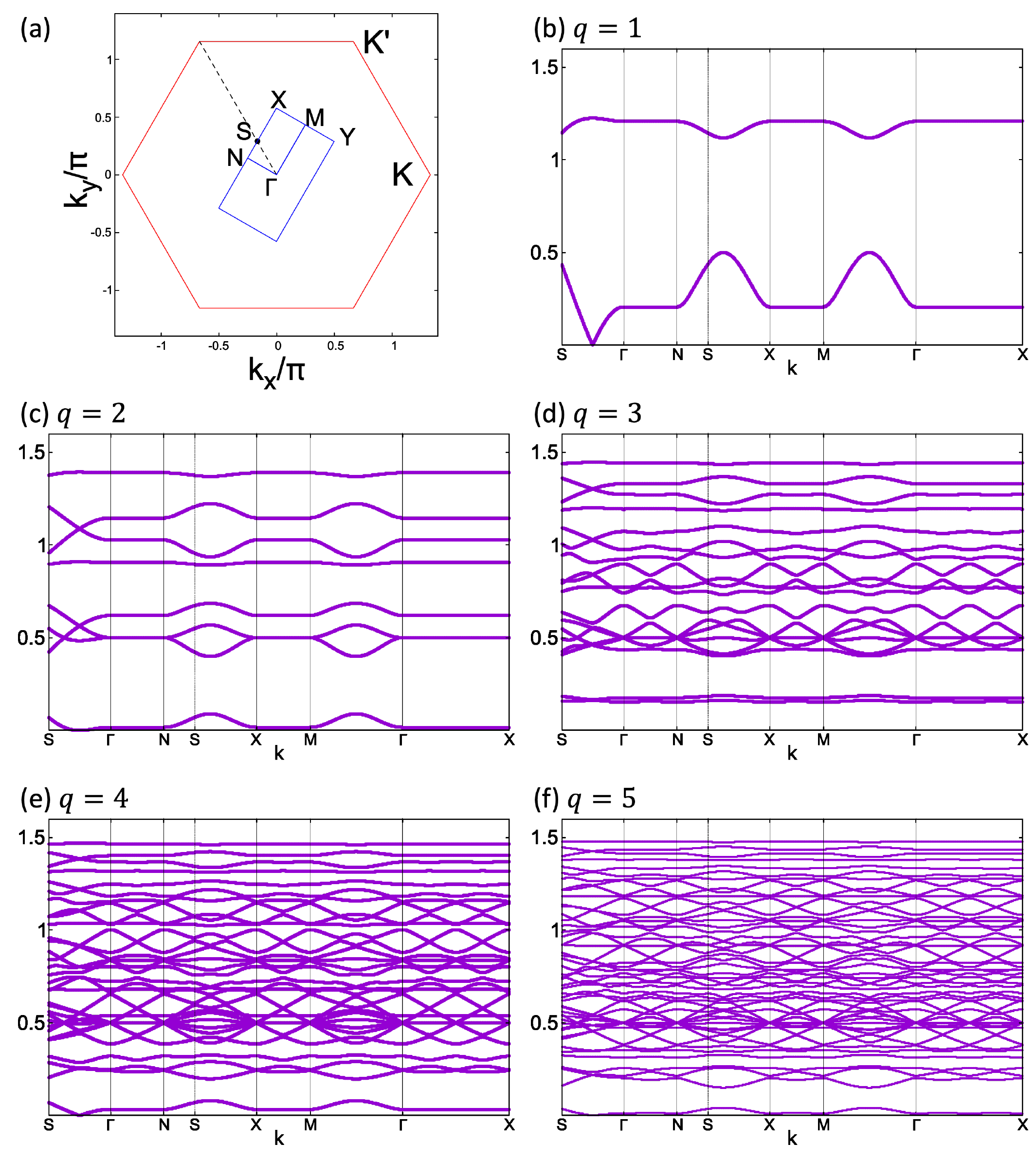}
  \caption{
    (a) reduced Brillouin zone in the Kitaev system with triangular flux configurations. 
    The dispersion relations for the flux configuations (b) $q=1$, (c) $q=2$, (d) $q=3$, (e) $q=4$ and (f) $q=5$.
  }
  \label{fig:dispersion}
\end{figure*}

\begin{figure}[htb]
  \includegraphics*[width = \linewidth]{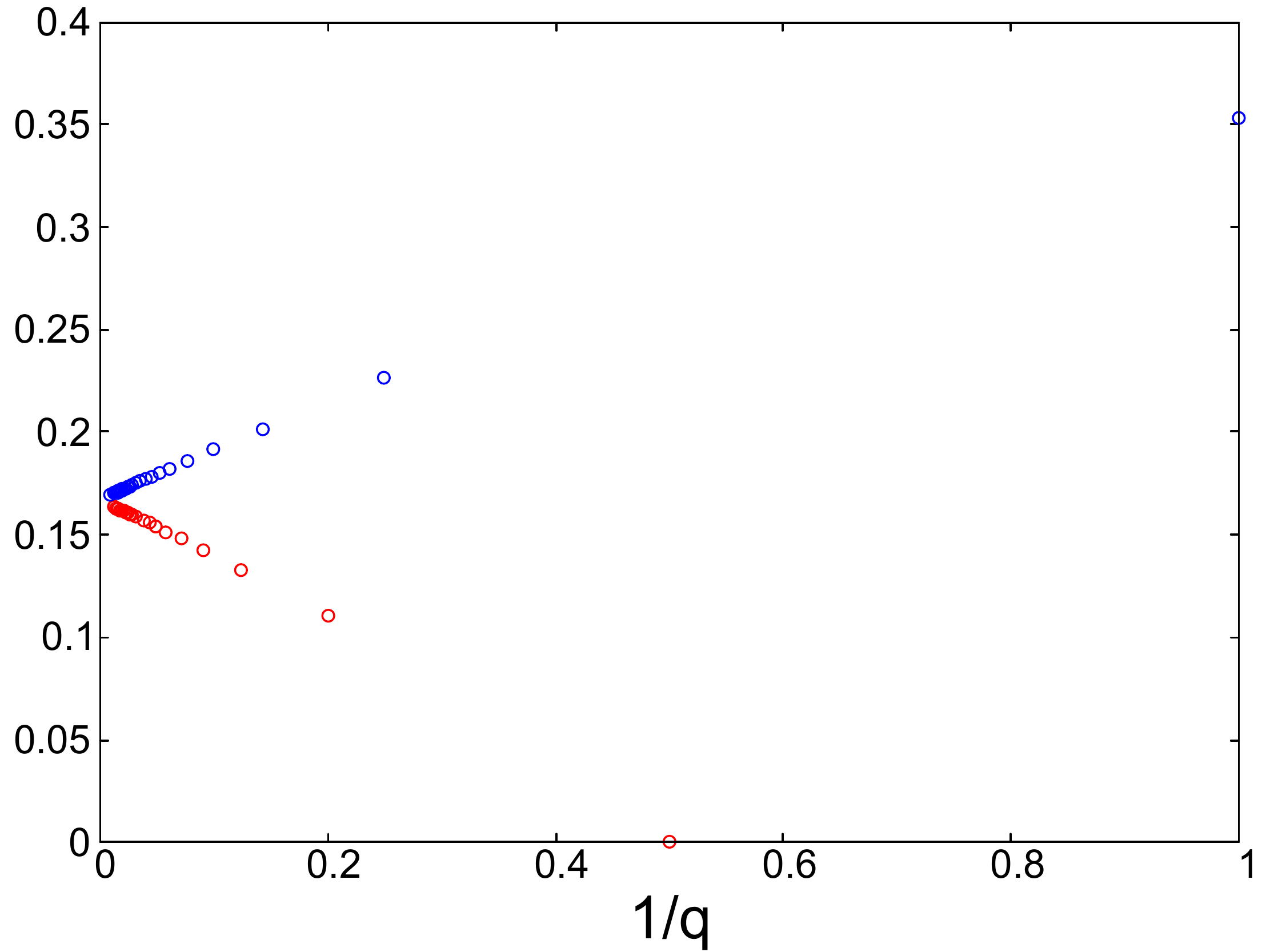}
  \caption{
    The magnitude of the gradient of the Dirac cone with respect to $1/q$.
    The blue represents $q\equiv1$ and the red represents $q\equiv2$.
    }
  \label{fig:gradient_q}
\end{figure}

Here, we discuss the Majorana excitations in the isotropic case $(J_x=J_y=J_z)$.
In the case, the system is essentially the same as the graphene system,
where the effect of the vortex configurations has been discussed~\cite{kamfor2011fate}.
The dispersion relations for $q=1, 2, 3, 4$, and 5 are shown in Fig.~\ref{fig:dispersion}.
It is found that the system is gapped in the case with $q=3$. 
On the other hand, in the others,
$q\neq 0$ (mod 3) and the system is gapless, as discussed in Sec.~\ref{sec:result}.
In fact, we find that the zero energy state always appears at the midpoint between S and $\Gamma$ points. 
Furthermore, the linear dispersion appears except for the configuration with $q=2$.
Figure~\ref{fig:gradient_q} shows the velocity of dispersion relation
calculated up to the configuration $q=100$,
which should be important in the Majorana-mediated transport~\cite{minakawa2020majorana,Taguchi_2021,Taguchi_2022}.
We clearly find that these can be divided into two groups $q=1$ or $2$ (mod 3),
which are shown as blue and red circles.
In the large $q$ limit, each velocity approaches $0.17J$,
which is different from that for the flux free case $v=\frac{\sqrt{3}}{4}J$.
Since the energy scale for the Majorana excitation originated from the flux configuration
should be tiny in the large $q$ case,
the Majorana excitations are almost described 
by the flux-free Kitaev model except for the lowest energy states.
Since the spin transport in the Kitaev model is mediated by the Majorana fermions
with finite energy, 
a small number of fluxes has little effect on the velocity of the spin transport.
The deital of the phase shift (Aharonov-Bohm effect) due to an isolated flux
has been discussed~\cite{Nasu2022}.

\bibliography{./refs}

\end{document}